\documentclass[prl,aps,twocolumn,showpacs]{revtex4}

\usepackage[dvips]{graphicx}
\usepackage{dcolumn}%
\usepackage{amsmath}%
\usepackage{amsfonts}%
\usepackage{amssymb}
 \usepackage[usenames]{color}
\usepackage{appendix}

\providecommand{\U}[1]{\protect\rule{.1in}{.1in}}
\newcommand{\be}{\begin{equation}}
\newcommand{\ee}{\end{equation}}
\newcommand{\bea}{\begin{eqnarray}}
\newcommand{\eea}{\end{eqnarray}}
\newcommand{\bt} {\begin{tabular}}
\newcommand{\et} {\end{tabular}}
\newcommand{\nn}{ \nonumber}
\newcommand{\ds}{\displaystyle}
\newcommand{\ba} {\begin{array}}
\newcommand{\ea} {\end{array}}
\begin{document}

\title{Heat currents in a two channel Marcus molecular junction}

\author{  Natalya A. Zimbovskaya {\footnote{Corresponding author: natalya.zimbovskaya@upr.edu}}}

\affiliation
{Department of Physics and Electronics, University of Puerto Rico-Humacao, CUH Station, Humacao, PR 00791, USA}

\begin{abstract}
We present a theoretical analysis of heat transport through a single-molecule junction with two possible transport channels for electrons where interactions between electrons on the molecule and phonons in the nuclear environment is strong and Marcus-type processes predominate in the electron transport. We show that  within the steady state regime the competition between transport channels may result in negative differential heat conductance and cooling of the  molecule environment. Also, we analyze the effect of a slowly driven molecule level (provided that another level is fixed) on the heat transport and power generated in the system.
\end{abstract}

\date{\today}
\maketitle

\subsection{I. Introduction} 

Presently, molecular electronics \cite{1,2,3,4,5} is a fast developing field providing a general platform to realize diverse atomic-scale devices. The basing building block for such devices is a single molecule junction (SMJ) that is a molecule linking two conducting (metallic/semiconductor) electrodes. Electron transfer through SMJs may be driven by electric forces and thermal gradients. A SMJ often operates being immersed in a dielectric solvent, and the solvent response may strongly affect electron transport through the molecule \cite{6,7,8}. In general, one may separate out two extreme limits for the electron transport through a junction. Within one limit the transport is nearly ballistic, and electron interactions with vibrational modes associated with the molecule as well as with thermalized phonons associated with its ambience may be treated as perturbations \cite{9,10,11}. Within another limit, the  effect of solvent environment is strong and electron transfer  may be viewed as a sequence of hops between the electrodes and the states on the molecular linker where a traveling electron may be transiently localized by distorting its close ambience. 

In the regime of strong electron-phonon interaction electron transport along molecules may be analyzed by using Marcus theory \cite{12,13,14} or its extensions \cite{15,16,17,18,19,20}. Marcus theory was repeatedly and successfully employed to study charge transport through molecules \cite{21,22,23,24,25,26,27}. In particular, it was shown that in redox molecular junctions whose operation involves reversible transitions between several oxidation states, influence of the molecular ambience may result in such interesting effects as charge current rectification and NDR \cite{23,24,27,28,29,30}. Heat transfer accompanying the charge transport in Marcus junctions was also studied \cite{15,16,31}.

Nevertheless, the analysis of heat conduction through Marcus SMJs is not completed so far, especially, in the case of redox junctions. In Sec.II of the present work we consider steady state heat currents through a SMJ with two transport channels within Marcus transport regime. We show that competition between the channels may result in negative differential heat conduction (NDHC) and solvent cooling. In Sec.III, we analyze the energy balance in this system assuming that one of the bridge level is slowly driven by an external force. We also discuss the irreversible work done on the system and the corresponding dissipated power.
Conclusions are presented in Sec.IV.

\subsection{II. Steady state heat currents in a two channel system}

As a model for the two channel bridge we choose a molecule with three states $|a>$, $|b>$ and $|c>$ accessible within the considered range of the bias voltage $V$.We assume that the states $|a>$ and $|c>$ are different charged states of the molecule, and the molecule is neutral being in the state $|b>$. Probabilities $P_a$, $P_b$ and $P_c$ for the molecule to be in these states at a certain moment $t$ ($P_a+P_b+P_c=1$) are given by kinetic equations \cite{27,32}:
\be
\frac{dP_a}{dt}=P_b\cdot k_{ba}-P_a\cdot k_{ab}                   \label{1}
\ee
\be
\frac{dP_b}{dt}=P_a\cdot k_{ab}+P_c\cdot k_{ac}-P_b\cdot\left(k_{ba}+k_{bc}\right)           \label{2}
\ee
\be
\frac{dP_c}{dt}=P_b\cdot k_{bc}-P_c\cdot k_{cb}                              \label{3}
\ee
Here, $k_{\alpha b}=k^{L}_{\alpha b}+k^{R}_{\alpha b}$, $\alpha=\{a,c\}$ and Marcus approximations for the transfer rates are given by \cite{12,13}:
\begin{align}
k_{\alpha b}^K = &\sqrt{\frac{\beta_s}{4\pi \lambda_{\alpha}}} \Gamma^{K}_{\alpha}\int_{-\infty}^\infty d \epsilon [1 - f_K (\beta_K, \epsilon)] 
\nn\\ & \times
\exp \left[-\frac{\beta_s}{4 \lambda_{\alpha}} (\epsilon + \lambda_{\alpha} - \epsilon_{\alpha})^2 \right]   \label{4},
\end{align}
\begin{align}
k_{b \alpha}^K = &\sqrt{\frac{\beta_s}{4\pi\lambda_{\alpha}}}\Gamma^{K}_{\alpha} 
\int_{-\infty}^\infty d \epsilon f_K (\beta_K, \epsilon)
\nn\\ &\times
\exp\left[- \frac{\beta_s}{4\lambda_{\alpha}} (\epsilon_\alpha + \lambda_{\alpha} - \epsilon)^2 \right],   \label{5}
\end{align}
Here, $K=\{L,R\}$, $\epsilon_\alpha=E_{\alpha}-E_b$ ($E_{\alpha}, E_b$ being the energies associated with molecular states $|\alpha>$ and $|b>$), $\lambda_{\alpha}$ are reorganization energies corresponding to $|\alpha> \to |b>$ and $|b>\to|\alpha>$ transitions, $\Gamma^{K}_{\alpha}$ are bare electron transfer rates between the molecular state $|\alpha>$ and the left/right electrode, $\beta_{K}=\ds\frac{1}{kT_K}$ and $\beta_s=\ds\frac{1}{kT_s}$ indicate the temperatures of the electrodes and that of the solvent, $k$ is the Boltzmann constant and $f_K(\beta_K,\epsilon)$ are Fermi distribution functions for the electrodes with chemical potentials $\mu_K$. As follows from these expressions transfer rates $k_{\alpha b}$ and $k_{b\alpha}$ refer to the electron removal from the molecule and injection into one of the relevant molecular states. Within our model we assume that only a single electron may be injected/removed to/from the states $|a>$ and $|c>$. States corresponding to a doubly charged molecule are supposed to be inaccessible within the bias voltage range. In further analysis we assume that $T_{L}=T_{R}=T_{s}$ and the bias voltage $V$ is symmetrically distributed between the electrodes: $\mu_{L,R}=\mu\pm \ds\frac{e}{2}V$ where the chemical potential $\mu$ is corresponding to an unbiased system. Also we assume that one of the states (for certainty we choose $|c>$) is asymmetrically coupled to the electrodes, that is $\Gamma_{a}^{L}=\Gamma_{a}^{R}=\Gamma_{c}^{L}\gg\Gamma_{c}^{R}$ and $\epsilon_{c}>\epsilon_a$

Steady state probabilities $P_{a}^0$, $P_{b}^0$ and $P_{c}^0$ may be computed from Eqs.(\ref{1})-(\ref{3}):
\be
P_{b}^0=\frac{1}{1+\ds\frac{k_{ba}}{k_{ab}}+\ds\frac{k_{bc}}{k_{cb}}}; \qquad  P_{a}^0=P_{b}^0\frac{k_{ba}}{k_{ab}}; \qquad  P_{c}^0=P_{b}^0\frac{k_{bc}}{k_{cb}}.            \label{6}                 
\ee
\begin{figure}[t] 
\begin{center}
\includegraphics[width=8cm,height=6cm]{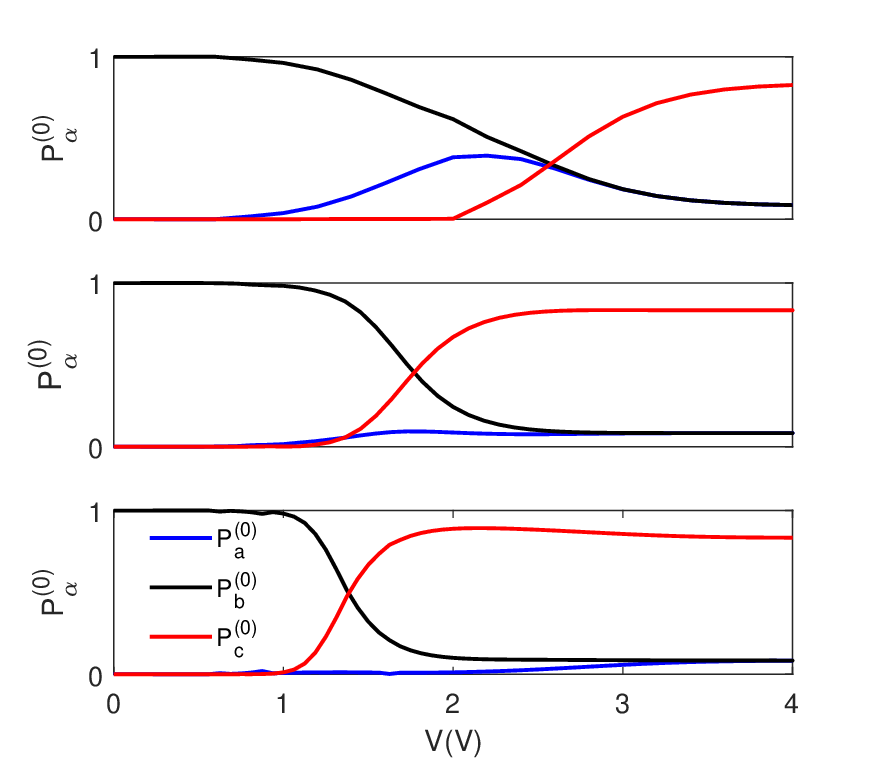} 
\caption{Steady state probabilities plotted for a two channel Marcus SMJ as functions of the bias voltage assuming that $T_{L}=T_{R}=T_{s}= 0.026$eV, $\Gamma_{a}^{L}=\Gamma_{a}^{R}=\Gamma_{c}^{L}=10\Gamma_{c}^{R}$, $\epsilon_{a}=0.2$eV, $\epsilon_{c}=0.6$eV and $\lambda_{a}=0.5$eV, $\lambda_{c}=0.6$eV (top panel); $\lambda_{a}=0.9$eV, $\lambda_{c}=0.5$eV (middle panel); $\lambda_{a}=1.2$eV, $\lambda_{c}=0.25$eV (bottom panel).
}
 \label{rateI}
\end{center}\end{figure}
and used to compute the steady state charge current $I_{s}$ given by the expression:
\be
\frac{I_{s}}{e}=k_{ba}^{L}P_{b}^0+k_{bc}^{L}P_{b}^0-k_{ab}^{L}P_{a}^0-k_{cb}^{L}P_{c}^0           \label{7}
\ee
which could be reduced to the form \cite{27}:
\be
\frac{I_s}{e}=\frac{\left(I_1\left(1+\ds\frac{k_{ba}}{k_{ab}}\right)+I_2\left(1+\ds\frac{k_{bc}}{k_{cb}}\right)\right)}{\left(1+\ds\frac{k_{ba}}{k_{ab}}+\ds\frac{k_{bc}}{k_{cb}}\right)}                       \label{8}
\ee
where
\be
I_1=\frac{k^{R}_{ab}k^{L}_{ba}-k^{L}_{ab}k^{R}_{ba}}{k_{ab}+k_{ba}};   \qquad  I_2=\frac{k^{R}_{cb}k^{L}_{bc}-k^{L}_{cb}k^{R}_{bc}}{k_{cb}+k_{bc}}  \label{9}
\ee
are currents flowing through the first and second transport channel, respectively. In the expression for $I_{s}$ each of these currents is multiplied by the probability that the corresponding channel is open.
In general, the steady state electron current cannot be reduced to the sum of $I_1$ and $I_2$. However, as we assumed that $\epsilon_{c}>\epsilon_{a}$,  at sufficiently low bias when $P^0_{c}=0$ the total current equals $I_1$, which is illustrated in the upper panel of Fig.1. 

As discussed in earlier works \cite{27,32} a two channel system may show NDR effect in current-voltage curves provided that the transmission channel accessible at a higher bias voltage has a blocking character, that is the system relatively easily switches to the state with the higher energy ( in our case from $|b>$ to $|c>$) but takes a significantly longer time to switch back. In the considered case this could be achieved when the coupling of the higher molecular orbital to the electrodes is asymmetric, namely $\Gamma_{c}^{R}\ll\Gamma_{c}^{L}$ and the system is stabilized by solvent reorganization occurring when the state $|c>$ becomes occupied. Competition between the transport channels results in decrease of the conduction at the threshold voltage corresponding to population of $|c>$ which is manifested as NDR. The NDR disappears when both $|a>$ and $|c>$ become occupied at the same voltage, as shown in the middle panel of Fig.1 and the transport channels are simultaneously opening. Finally, it may happen that the reorganization energy $\lambda_a$ significantly exceeds $\lambda_c$ and the channel associated with the higher energy $\epsilon_c$ opens up at lower bias voltage, as illustrated in the bottom panel of Fig.1. This inverted order of accessibility of the transport channels does not lead to NDR in current-voltage curves \cite{27} but may significantly affect heat currents, as shown below.

Each electron hop between the molecule and an electrode is accompanied by heat production in both electrodes and solvent environment of the molecule originating from their relaxation. We denote the heat produced in the solvent as $Q_s$ and that produced in the electrodes as $Q_e$. Specifically, $Q^{K}_{s, \alpha b}$ and $Q^{K}_{s,b\alpha}$ are heat changes in the solvent when an electron hops to (from) $K$ electrode from (to) the molecule state $|\alpha>$. Within Marcus approach these heats may be written in the form similar to that used in earlier works \cite{16}:
\begin{align}
Q_{s,\alpha b}^{K} = & \frac{\Gamma^{K}_{\alpha}}{k_{\alpha b}^{K}} \sqrt{\frac{\beta_s}{4 \pi\lambda_{\alpha}}} \int_{-\infty}^{\infty} d \epsilon \big[1 - f_{K} (\beta_{K},\epsilon) \big] (\epsilon_{\alpha}-\epsilon )  
\nn\\ & \times
\exp \left[- \frac{\beta_s}{4\lambda_{\alpha}} (\lambda_{\alpha} - \epsilon_{\alpha}+ \epsilon)^2 \right].   \label{10}
\end{align}
and
\begin{align}
Q_{s,b \alpha}^{K} = & \frac{\Gamma^{K}_{\alpha}}{k_{b\alpha}^{K}} \sqrt{\frac{\beta_s}{4 \pi \lambda_{\alpha}}} \int_{-\infty}^{\infty} d \epsilon f_{K} (\beta_{K},\epsilon) ( \epsilon - \epsilon_{\alpha}) 
\nn \\ & \times
\exp \left[- \frac{\beta_s}{4\lambda_{\alpha}} ( \epsilon_{\alpha} + \lambda_{\alpha} - \epsilon)^2 \right].   \label{11}
\end{align}

Heats $Q^{K}_{e,\alpha b}$ and $Q^{K}_{e,b\alpha }$ generated in the electrode $K$ when an electron leaves (enters) $|\alpha>$ state on the molecule and arrives to (leaves from) this electrode  may be approximated by the following expressions:
\begin{align}
Q_{e,\alpha b}^{K} = & \frac{\Gamma^{K}_{\alpha}}{k_{\alpha b}^{K}} \sqrt{\frac{\beta_s}{4 \pi\lambda_{\alpha}}} \int_{-\infty}^{\infty} d \epsilon \big[1 - f_{K} (\beta_{K},\epsilon) \big] (\epsilon -\mu_{K})  
\nn\\ & \times
\exp \left[- \frac{\beta_s}{4\lambda_{\alpha}} (\lambda_{\alpha} - \epsilon_{\alpha} + \epsilon)^2 \right].   \label{12}
\end{align}
and
\begin{align}
Q_{e,b \alpha}^{K} = & \frac{\Gamma^{K}_{\alpha}}{k_{b\alpha}^{K}} \sqrt{\frac{\beta_s}{4 \pi\lambda_{\alpha}}} \int_{-\infty}^{\infty} d \epsilon f_{K} (\beta_{K},\epsilon) (\mu_{K}- \epsilon ) 
\nn\\ & \times
\exp \left[- \frac{\beta_s}{4\lambda_{\alpha}} ( \epsilon_{\alpha} + \lambda_{\alpha} - \epsilon)^2 \right].   \label{13}
\end{align}
In Eqs.(\ref{10})-(\ref{13}) the coupling parameters $\Gamma^{K}_{\alpha}$ are supposed to be independent on the energy, and they are treated as constants.

The corresponding  heat change rates (heat currents) in the solvent ($J_s=\dot{Q}_s$) and electrodes ($J^{K}_e=\dot{Q}^{K}_e$) are:
\begin{align}
J_s=& P^0_{a}(k_{ab}^{L}Q_{s,ab}^{L}+k_{ab}^{R}Q_{s,ab}^{R})+P^0_{c}(k_{cb}^{L}Q_{s,cb}^{L}+k_{cb}^{R}Q_{s,cb}^{R})
\nn\\ &
+P^0_{b}(k_{ba}^{L}Q_{s,ba}^{L}+k_{ba}^{R}Q_{s,ba}^{R}+k_{bc}^{L}Q_{s,bc}^{L}+k_{bc}^{R}Q_{s,bc}^{R})        \label{14}
\end{align}   
 and:
\begin{align}
J^{K}_{e}= & P^0_{a}k_{ab}^{K}Q_{e,ab}^{K}+P^0_{b}k_{ba}^{K}Q_{e,ba}^{K}
\nn\\ &
+P^0_{c}k_{cb}^{K}Q_{e,cb}^{K}+P^0_{b}k_{bc}^{K}Q_{e,bc}^{K}                               \label{15}
\end{align}
Summing up all heat currents and using Eqs.(\ref{4}),(\ref{5}) as well as Eqs.(\ref{10}) -(\ref{13}) we may show that Eqs.(\ref{14}) and (\ref{15}) imply that:
\be
J^{L}_{e}+J^{R}_{e}+J_{s}=(\mu_{L}-\mu_{R})\frac{I_{s}}{e}                                 \label{16}
\ee
thus conforming the balance between the power given to the system by applying the bias voltage ($\mu_{L}-\mu_{R}=eV$) and the heat currents deposited into the electrodes and the solvent.

Negative differential heat conductance appears in the $J_{s}$ dependencies on the applied bias voltage under the same conditions as NDR in the charge current
\begin{figure}[t] 
\begin{center}
\includegraphics[width=6cm,height=5cm]{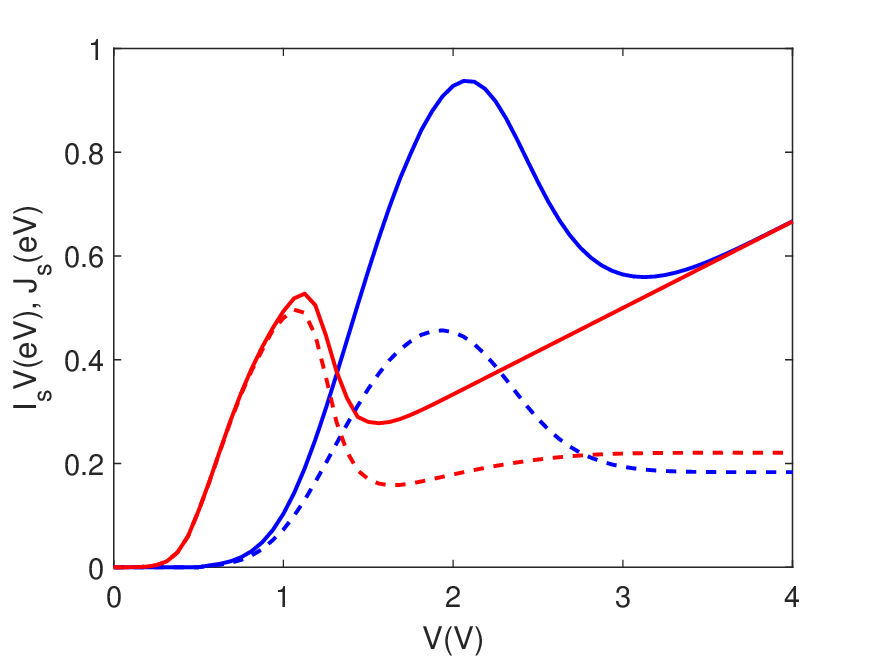} 
\includegraphics[width=6cm,height=5cm]{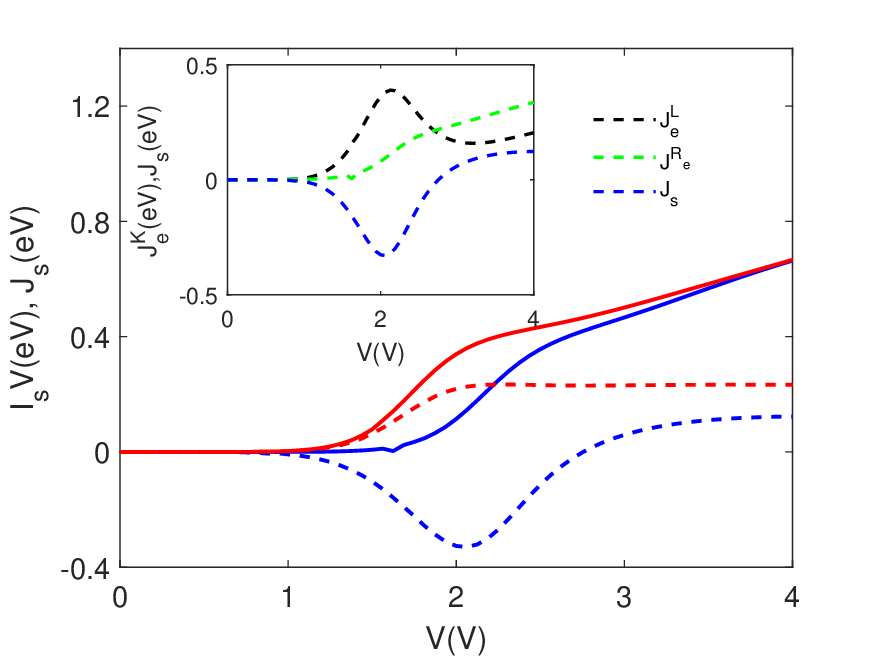}
\caption{The power $I_{s}V$ given to the biased system (solid lines) and the heat current $J_s$ flowing to the solvent (dashed lines) as functions of the bias voltage plotted at $T_{L}=T_{R}=T_{s}= 0.026$eV, $\Gamma_{a}^{L}=\Gamma_{a}^{R}=\Gamma_{c}^{L}=10\Gamma_{c}^{R}$, $\epsilon_{a}=0.2$eV, $\epsilon_{c}=0.6$eV. Left panel: $\lambda_{a}=\lambda_{c}$=0.1eV (red lines) and $\lambda_{a}=0.5$eV, $\lambda_{c}=0.6$eV (blue lines). Right panel: $\lambda_{a}=0.9$eV, $\lambda_{c}=0.5$eV (red lines) and $\lambda_{a}=1.2$eV, $\lambda_{c}=0.25$eV (blue lines). Inset shows $J^{K}_{e}$ and $J_{s}$ individually displayed as functions of $V$. 
}
 \label{rateI}
\end{center}\end{figure}
 versus voltage curves. This is displayed in Fig.2 (left panel). Unlike the heat current into the solvent the currents into electrodes do not show NDHC. Note that the peak in the $J_{s} -V$ curve shifts to the greater bias voltage values when reorganization energies increase. This happens because a stronger bias is required to overcome the Frank -Condon blockade \cite{33,34} originating from electron interaction with solvent phonons. Note that the ratio $\ds\frac{J_s}{I_{s}V}$ which represents the part of available energy deposited into the solvent takes on values close to $1$ when the bias voltage is moderate and the reorganization energies $\lambda_{\alpha}$ are low. For higher $\lambda_{\alpha}$ this ratio remains significantly smaller. This result seems somewhat surprising for the heat current into the solvent should vanish for zero reorganization energies. However, we must keep in mind that at low bias reorganization processes in the solvent  may play a significant part even at low (but nonzero) values of $\lambda_{\alpha}$.

As the difference between $\lambda_{a}$ and $\lambda_{c}$ ($\lambda_{a}>\lambda_{c}$ increases, the two transport channels become accessible nearly simultaneously, and the NDHC effect is fading away along with the NDR. This is illustrated in the right panel of Fig.2 (see red lines). However, an interesting behavior of heat currents appear at sufficiently large difference $\lambda_{a}-\lambda_{c}$ which ensures the inversion of the order in which the channels are accessed. In this situation, at a moderate bias one of the heat currents to the electrodes (in the considered case when $\epsilon_{a}<\epsilon_{c}$ and $\Gamma_{c}^{L}\gg\Gamma_{c}^{R}$ it should be $J^{L}_{e}$) strongly exceeds another one. At the same time, the heat current into the solvent takes on negative values indicating the solvent cooling. This could occur as a result of interplay between the electric driving forces and forces appearing due to solvent reorganization provided that the transport channel which opens up at lower bias voltage has a blocking character. Note that the solvent cooling may be caused by different mechanisms, as was shown for SMJs with low electron-phonon interactions \cite{35,36,37}.

\subsection{III. Heat currents and work in a driven junction.}

Now, we turn to the analysis of energy currents in a driven two channel junction. The driving is modeled by time dependence of one of the molecule states energy. For certainty, we choose the state $|a>$ as a driven state thus assuming that $\epsilon_{a}$ depends on time whereas $\epsilon_{c}$ remains fixed. The driving of $\epsilon_{a}$ may be achieved by varying the corresponding gate potential. Transport properties of driven junctions with negligible electron-phonon interactions and a single transport channel were studied in several works \cite{38,39,40,41,42,43,44,45,46}. The model considered here includes strong coupling to the phonon environment at the cost of treating this coupling semiclassically and assuming weak coupling between molecule and electrodes. Similar model was used to study heat currents and work done in a driven junction with a single transport channel were also \cite{31}.

We assume that the energy $\epsilon_{a}$ is varying slowly, that is $\dot{\epsilon}_{a}$ is small compared to $kT_{s}\Gamma^{K}_{\alpha}$ and $\ds\frac{(\Gamma^{K})^2}{h}$. Then we present the population probabilities as sums of their steady state values and time dependent corrections:
\be
P_{a}(t)=P_{a}^{0}(\epsilon_{a})-G_{a}(t);    \qquad  P_{c}(t)=P_{c}^{0}(\epsilon_{c})-G_{c}(t);       \label{17}
\ee
where the corrections obey the relationship: $G_{a}+G_{b}+G_{c}=0$. As shown in an earlier work \cite{31} these corrections may be expanded in powers of $\dot\epsilon_{a}$. Here, we restrict our consideration by first order corrections linear in $\dot\epsilon_{a}$ which is justified for the case of quasistatic processes. Then Using Eqs.(\ref{1}) -(\ref{3}) we get the following expressions for the these corrections:
\be
G_a=\dot\epsilon_{a}\frac{\partial P^{0}_a}{\partial\epsilon_a}\frac{k_{bc}+k{cb}}{(k_{ab}+k_{ba})(k_{cb}+k_{bc})-k_{bc}k_{ba}} ;    \label{18}
\ee
\be
G_b=-G_a\frac{k_{cb}}{k_{cb}+k_{bc}};            \qquad   G_c=-G_a\frac{k_{bc}}{k_{cb}+k_{bc}}.                    \label{19}
\ee 
Electronic currents now differ from their steady state values and acquire corrections proportional to $G_a$:
\be
 I_{L}=I_{s}+e\cdot G_a\left(k^{L}_{ab}+\frac{k_{cb}k^{L}_{ba}}{k_{bc}+k_{cb}}+I_2 \right)\equiv I_{s}+I^{(1)}_L                   \label{20}
\ee
\be
I_{R}=I_{s}+e\cdot G^{(1)}_a\left(k^{R}_{ab}+\frac{k_{cb}k^{L}_{ba}}{k_{bc}+k_{cb}}-I_2 \right)\equiv I_{s}+I^{(1)}_R                    \label{21}
\ee 
\begin{figure}[t] 
\begin{center}
\includegraphics[width=6cm,height=5cm]{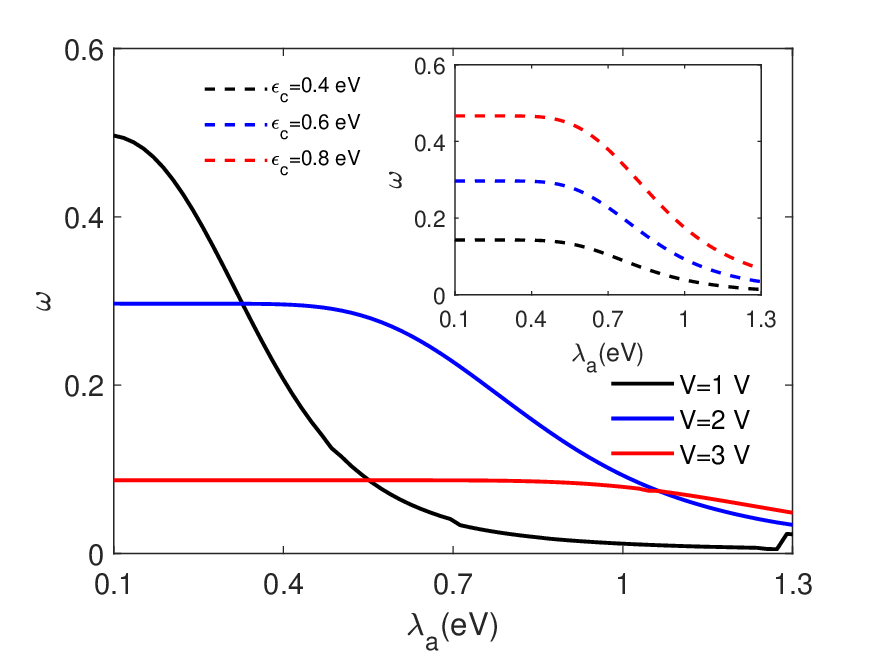} 
\includegraphics[width=6cm,height=5cm]{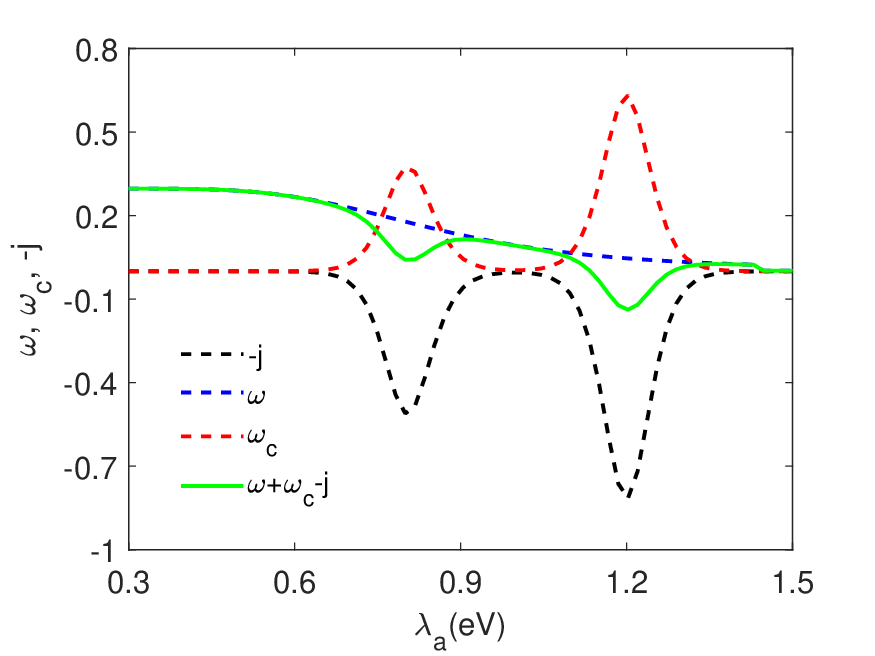}
\caption{Left panel: The reduced reversible power $\omega=\ds\frac{W}{\dot\epsilon_{a}}$ as a function of $\lambda_{a}$ displayed for several values of the bias voltage at the fixed energies $\epsilon_{c}$ and $\lambda_{c}$. Inset shows $\omega$ versus $\lambda_{a}$  plotted at $V=2$V and different values of $\epsilon_{c}$. Right panel: Dependencies of the reduced reversible power $\omega$, the rate of chemical work $\omega_{c}=\ds\frac{W_{chem}}{\dot\epsilon_{a}}$ and the total heat current $j=\ds\frac{J_{tot}}{\dot\epsilon_{a}}$ on $\lambda_{a}$. The green line is corresponding to $\dot{E}_{M}$. Curves are plotted assuming that $T_{L}=T_{R}=T_{s}= 0.026$eV, $\Gamma_{a}^{L}=\Gamma_{a}^{R}=\Gamma_{c}^{L}=10\Gamma_{c}^{R}$, $\epsilon_{a}=0.2$eV, $\epsilon_{c}=0.6$eV (right panel), $\lambda_{c}=0.6$eV. 
}
 \label{rateI}
\end{center}\end{figure}

Similarly, corrections proportional $G_a$ appear in the expressions for heat currents. Introducing the total heat current $J_{tot}=J^{L}_{e}+J^{R}_{e}+J_s$ we may present it in the form $J_{tot}=J^{(0)}+J^{(1)}_{tot}$ where $J^{(0)}$ is the steady state heat current which is equal to the sum of currents given by Eqs.(\ref{14}) and (\ref{15}) and the correction $J^{(1)}_{tot}$ is given by:
\be
J^{(1)}_{tot}=\mu_{L}I^{(1)}_L+\mu_{R}I^{(1)}_R-\dot\epsilon_{a}\epsilon_{a}\frac{\partial P^{(0)}_{a}}{\partial \epsilon_{a}}      \label{22}
\ee
To better clarify the meaning of this expression we rearrange the second term as $-\ds\dot\epsilon_{a}\epsilon_{a}\frac{\partial P^{(0)}_{a}}{\partial \epsilon_{a}}=\dot\epsilon_{a}P^{(0)}_{a}-\ds\frac{d}{dt}\left(\epsilon_{a}P^{(0)}_{a}\right)$. Then Eq.(\ref{22}) could be reduced to the form:
\be
\frac{d}{dt}\left(\epsilon_{a}P^{(0)}_{a}\right)=\dot\epsilon_{a}P^{(0)}_{a}+\mu_{L}I^{(1)}_L+\mu_{R}I^{(1)}_R-J^{(1)}_{tot}          \label{23}
\ee
which is an analog of the corresponding result derived for a single channel junction \cite{31}. This expression confirms the first law of thermodynamics written for a quasistatic process. On the left side we have the rate of change of the molecular energy  $\dot{E}_{M}=\ds\frac{d}{dt}\left(\epsilon_{a}P^{(0)}_{a}\right)$ caused by the driving of the lower molecular level. It is equal to the sum of the reversible power $W=\dot\epsilon_{a}P^{(0)}_{a}$, the rate of chemical work $W_{chem}=\dot\epsilon_{a}\ds\frac{\partial P^{(0)}}{\partial\epsilon_{a}}(\mu_{L}I^{(1)}_{L}+\mu_{R}I^{(1)}_{R})$ and the heat current coming from  the environment $-J^{(1)}_{tot}$.

All these terms strongly depend on the reorganization energy $\lambda_{a}$, as illustrated in Fig.3 where we show the behavior of reduced entities $\omega=\ds\frac{W}{\dot\epsilon_{a}}$, $\omega_{c}=\ds\frac{W_{chem}}{\dot\epsilon_{a}}$ and $j=\ds\frac{J_{tot}}{\dot\epsilon_{a}}$ assuming for certainty that $\dot\epsilon_{a}>0$. One observes that the reduced power $\omega$ which remains nearly constant at sufficiently small $\lambda_{a}$ noticeably decreases at higher values of the reorganization energy. This behavior could be caused by the Frank-Condon blockade. At each fixed bias, the blockade is lifted at sufficiently low reorganization energy which makes $\omega$ only weakly dependent on the latter. At higher $\lambda_{a}$ the Coulomb blockade emerges reducing molecular coupling to electrodes and, consequently, the reversible power. Also, the work done to drive one of the molecular levels with the energy $\epsilon_{a}$ is affected by the fixed level position. This is demonstrated in the inset.

To better understand how much each term in Eq.(\ref{23}) contributes to the change of the molecular energy, we separately plot them as functions of $\lambda_{a}$ in he right panel of Fig.3 at a fixed bias voltage $V$ and energy $\epsilon_{c}$. As shown in this figure, $\dot {E}_{M}$ practically coincides with the power term everywhere, except vicinities of the points $\lambda_{a}=\ds\frac{1}{2}V\pm\epsilon_{a}$. These points indicate the opening and closing of the lower transport channel as it crosses boundaries of the conduction window determined by the voltage value. Near these points both heat currents and chemical work strongly contribute to $\dot {E}_{M}$ but their contributions to a significant extent counterbalance each other, so the total effect remains rather moderate. To further elucidate the effect of the higher level transport channel on transport properties of a two channel system we study the behavior of $\omega$, $\omega_{c}$ and $j$ as functions of the energy $\epsilon_{c}$ at fixed $\epsilon_{a}$, $\lambda_{a}$ and $V$. Note that the chosen values of the latter three parameters indicate the opening of the transport channel associated with $\epsilon_{a}$. The results are displayed in Fig.4 (left panel). At moderate difference between $\epsilon_{c}$ and $\epsilon_{a}$ all contributions to  $\dot {E}_{M}$ remain independent on $\epsilon_{c}$. However, when $\epsilon_c$ approaches the value corresponding to the inversion of the order in which the two channels become accessible, all contributions to $\dot {E}_{M}$ significantly increase in magnitudes, and this entity itself increases before anew becoming a constant at greater values of $\epsilon_{c}$. Note that at other conditions, unfavorable for opening/closing the first transport channel, $\dot {E}_{M}$ is dominated by the reversible power $W$, contributions from other terms in Eq.({\ref{23}) being negligible, as demonstrated in the inset.
\begin{figure}[t] 
\begin{center}
\includegraphics[width=6cm,height=5cm]{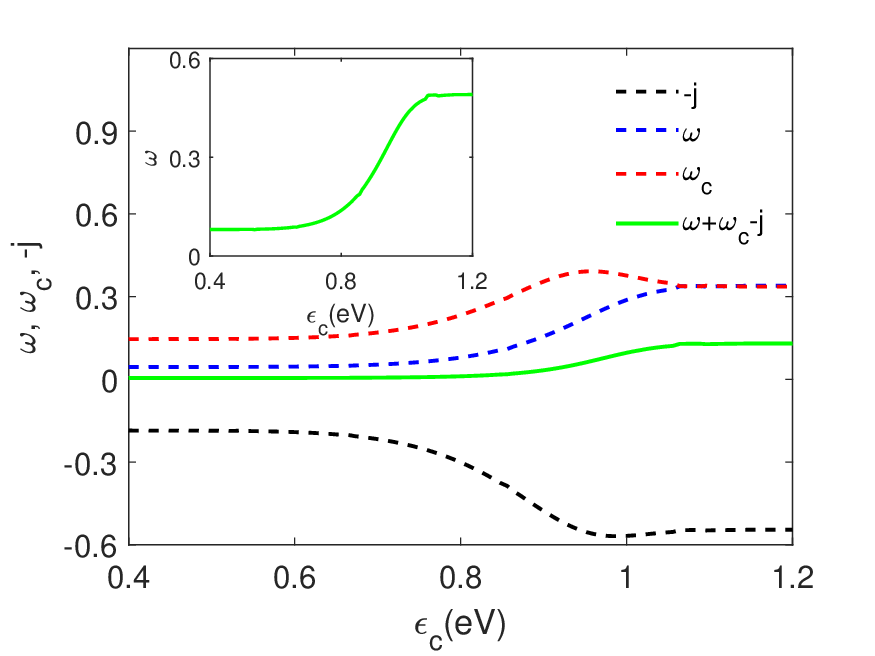} 
\includegraphics[width=6cm,height=5cm]{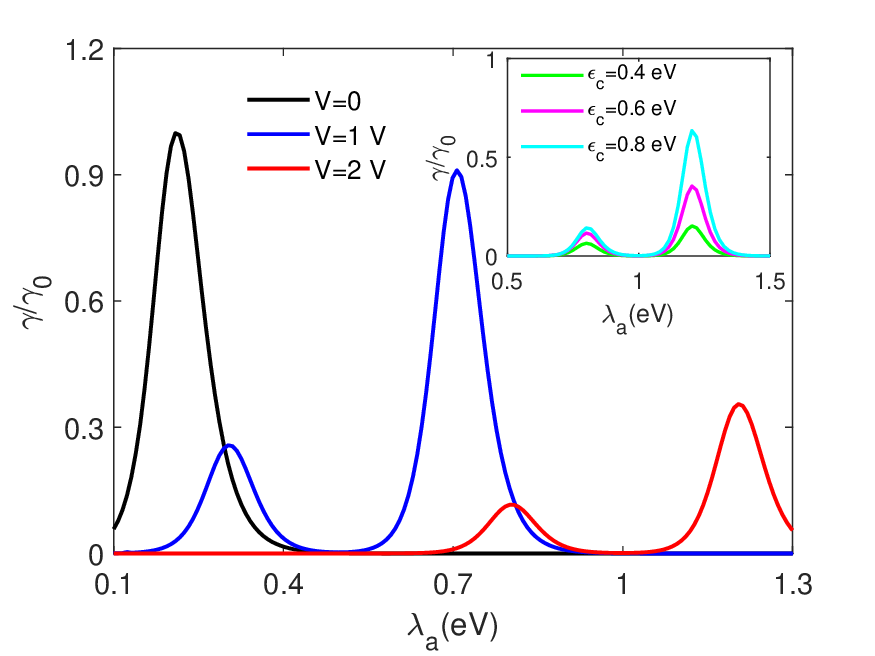}
\caption{Left panel:  Dependencies of $\omega$, $\omega_{c}$ and $-j$ on $\epsilon_{c}$ plotted at $V=2$V, $\epsilon_{a}=0.2$eV, $\lambda_{a}=0.8$eV, $\lambda_{c}=0.2$ eV. The green line is corresponding to $\dot{E}_{M}$. Right panel: Friction coefficient $\gamma$ as a function of the reorganization energy $\lambda_{a}$. Curves are plotted assuming that $T_{L}=T_{R}=T_{s}= 0.026$eV, $\Gamma_{a}^{L}=\Gamma_{a}^{R}=\Gamma_{c}^{L}=10\Gamma_{c}^{R}$, $\epsilon_{a}=0.2$eV, $\epsilon_{c}=0.6$eV. (left panel) $\lambda_{c}=0.6$eV.  
}
 \label{rateI}
\end{center}\end{figure}

When the system operates beyond the quasistatic regime, the work done to drive the molecular level acquires an irreversible contribution $\Delta\dot{W}$ which may be approximated as$\dot\epsilon_{a} G_{a}$. This correction is associated with energy dissipation accompanying the level driving. Following Refs. \cite{38,47,48} we may present this irreversible power which is quadratic in $\dot\epsilon_{a}$ in the form $\Delta\dot{W}=\gamma\dot\epsilon_{a}^{2}$ where $\gamma$ is the friction coefficient. Using the expression (\ref{18}) for $G_{a}$ we get:
\be
\gamma=-\frac{\partial P^{(0)}_{a}}{\partial\epsilon_{a}}\frac{k_{cb}+k_{bc}}{(k_{ab}+k_{ba})(k_{cb}+k_{bc})-k_{bc}k_{ba}}            \label{24}
\ee
The behavior of the friction coefficient is presented in the right panel of Fig.4. One observes that in the case of an unbiased junction the friction coefficient shows a single peak which emerges when the difference $\epsilon_{a}-\lambda_{a}=0$ indicating that the transport channel associated with the lower energy $\epsilon_{a}$ crosses the chemical potential of electrodes $\mu_{L}=\mu_{R}=\mu$. We denote this peak height as $\gamma_0$. When the system is biased the peak becomes split in two indicating the two crossings by the transport channel of the boundaries of the conduction window. The heights of these peaks are smaller than $\gamma_0$ and they decrease as the bias strengthens. Also, at a stronger bias, the peaks are situated at greater values of $\lambda_a$. Such behavior of the friction coefficient is caused by the interplay between the electric driving force and the force appearing as a result the direct effect of the coupling of the molecule to the solvent. The presence of the second transport channel affects the behavior of $\gamma$ as well. An example of the second channel effect on the friction is shown in the inset. It is demonstrated that the friction becomes stronger as $\epsilon_{c}$ enhances.

\subsection {IV. Conclusions}

In the present work we have studied heat currents in single molecule junctions modeled by two level molecule connecting free electron metal electrodes and immersed in a solvent which may strongly affect electron transport along the molecule due to energy exchange between the molecule and the solvent accompanying the latter. Charge transfer kinetics was described by Marcus electron transfer theory. 

It was shown that within the steady state regime the competition between two transport channel may result in the weakening of the heat current flowing into the solvent which accompanies the strengthening of the bias voltage, provided that one of the channels is asymmetrically coupled to the electrodes. This effect is an analog of NDR which may appear in a two channel system \cite{27}. This effect fades away when both transport channels are  simultaneously accessible. A cooling of the solvent is predicted at sufficiently large differences in reorganization energies corresponding to the transport channels ensuring the inversion of the order in which the channels could be assessed in conformity with the associated energies.

Also, the heat currents and power produced by slow moving one of the electron levels (for certainty we choose the level corresponding to the state $|a>$) across a potential bias are studied. Accounting for the total molecular energy rate and its heat, work and chemical components computed up to the terms linear in $\dot\epsilon_{a}$, it is established that the energy conservation is satisfied in the considered system when the driving is a quasistatic process. Separate studies of the total molecular energy rate components behavior show that the chemical work and heat coming from the electrodes and from the solvent take a significant part in $\dot{E}_{M}$ only when the driven level crosses the boundaries of the conduction window determined by the bias voltage. Otherwise, these contributions remain negligible.

Beyond the quasistatic limit the power acquires an irreversible contribution associated with the friction associated with the friction appearing due to the electron exchange between the molecule and electrodes. It is shown that the friction coefficient in the considered system depends on the energies $\epsilon_{a}$ and $\epsilon_{c}$ and is affected by solvent reorganization accompanying the level driving.

The processes of energy conversion and heat transfer in nanoscale systems continue to attract interest of the research community. We believe that the present results may be useful for better understanding of these processes.

\subsection{Declaration of competing interest}

Authors declare that they have no competing financial interests or personal relationships which could influence the work reported in this paper.

\subsection{Data availability statement}

Data sharing is not applicable as no data are created in this study.

\subsection{Acknowledgments}

The present work was supported by the U.S National Science Foundation (DMR-PREM 2122102).

\end{document}